\documentclass[sn-mathphys,Numbered]{sn-jnl}

\usepackage{graphicx}%
\usepackage{multirow}%
\usepackage{amsmath,amssymb,amsfonts}%
\usepackage{amsthm}%
\usepackage{mathrsfs}%
\usepackage[title]{appendix}%
\usepackage{xcolor}%
\usepackage{textcomp}%
\usepackage{manyfoot}%
\usepackage{booktabs}%
\usepackage{algorithm}%
\usepackage{algorithmicx}%
\usepackage{algpseudocode}%
\usepackage{listings}%

\theoremstyle{thmstyleone}%
\theoremstyle{thmstyletwo}%

\theoremstyle{thmstylethree}%

\def\bra{\langle}
\def\ket{\rangle}

\def\calN{{\cal N}}

\def\pii[#1]{{1\over (2\pi)^{#1}}}

\newcommand{\xbld}[1]{\mbox{\boldmath $#1$}}

\def\vecr{\xbld{r}}

\begin{document}

\title[The impact of quark many-body effects on exotic hadrons]{The impact of quark many-body effects on exotic hadrons}


\author*[1,4,5]{\fnm{Sachiko} \sur{Takeuchi}}\email{s.takeuchi@jcsw.ac.jp}

\author[2,5,6]{\fnm{Makoto} \sur{Takizawa}}\email{takizawa@ac.shoyaku.ac.jp}
\equalcont{These authors contributed equally to this work.}

\author[3,4,7]{\fnm{Yasuhiro} \sur{Yamaguchi}}\email{yamaguchi@hken.phys.nagoya-u.ac.jp}
\equalcont{These authors contributed equally to this work.}

\author[4]{\fnm{Atsushi} \sur{Hosaka}}\email{hosaka@rcnp.osaka-u.ac.jp}
\equalcont{These authors contributed equally to this work.}

\affil*[1]{
\orgname{Japan College of Social Work}, \orgaddress{\street{Takeoka 3-1-30}, \city{Kiyose}, \postcode{204-8555}, \state{Tokyo}, \country{Japan}}}

\affil[2]{
\orgname{Showa Pharmaceutical University}, \orgaddress{\street{Higashi Tamagawa Gakuen 3-3165}, \city{Machida}, \postcode{194-8543}, \state{Tokyo}, \country{Japan}}}

\affil[3]{\orgdiv{Department of Physics, Faculty of Science}, \orgname{Nagoya University}, \orgaddress{\street{Chikusa-ku Furocho}, \city{Nagoya}, \postcode{464-8601}, \state{Aichi}, \country{Japan}}}

\affil[4]{\orgdiv{Research Center for Nuclear Physics (RCNP)}, \orgname{Osaka University}, \orgaddress{\street{Mihogaoka 10-1}, \city{Ibaraki}, \postcode{567-0047}, \state{Osaka}, \country{Japan}}}

\affil[5]{\orgdiv{Nishina Center}, \orgname{RIKEN}, \orgaddress{\street{Hirosawa}, \city{Wako}, \postcode{351-0198}, \state{Saitama}, \country{Japan}}}


\affil[6]{\orgdiv{J-PARC Branch, KEK Theory Center}, \orgname{KEK}, \orgaddress{\street{Shirakata 203-1}, \city{Tokai}, \postcode{319-11006}, \state{Ibaraki}, \country{Japan}}}

\affil[7]{Kobayashi-Maskawa Institute for the Origin of Particles and the Universe, Nagoya University, Nagoya, 464-8602, Japan}


\abstract{We investigate the exotic hadrons consisting of two light quarks and two heavy antiquarks, $(q\bar Q)$-$(q\bar Q)$. 
The spin-dependent term between quarks is known to give an attraction to the $ud$ spin-0 component in
 the isospin-0 $u\bar c d\bar c$ system, $T_{cc}$. 
However, the said component also gets a repulsion from the partial Pauli-blocking.
By the dynamical calculation with a simplified quark model, 
we discuss that the competition of the two effects leads to a shallow bound state for $T_{cc}$,
which is preferred from the experiment,
 and
a deep bound state for $T_{bb}$.
}




\maketitle

\section{Introduction}\label{sec1}

In 2022, the $T_{cc}$, a doubly charmed meson with $I(J^P)=0(1^+)$, was found near the threshold of $D^{*+}D^0$ by LHCb\cite{LHCb:2021auc}.
It is understood as an exotic state whose quark configuration is  $\bar u \bar d cc$.
Since the energy of the observed peak is just below the threshold with a binding energy of $\sim 0.3$ MeV,
it is considered a very shallow bound state of the $D$ and $D^*$ mesons.

Many theoretical works have been proposed to understand this state.
For example, extensive calculation by a quark model with and without the one-pion exchange (OPEP) \cite{Meng:2020knc,Meng:2023for}
or the chiral quark model with color-magnetic interaction (CMI) \cite{He:2023ucd},
or that with OPEP\cite{Sakai:2023syt} are reported after the experiments.
A recent review of the exotic hadrons is, for example, in ref.\ \cite{Chen:2022asf}. 

In this work, we investigate the exotic hadrons that consist of two light quarks $qq$
and two heavy antiquarks $\bar Q\bar Q$, namely, 
$(q\bar Q)$-$(q\bar Q)$ with $0(1^+)$
employing a simple quark model. 
The quark model successfully explains the features of most of the low-lying single hadrons. 
The hadron interactions are also well-reproduced by the quark model picture;
they are consistent with the known experiments, 
and the channel dependence of the short-range part of two-baryon
interaction corresponds to that given by the Lattice QCD \cite{Inoue:2018axd}.

The short-range part of such hadron interactions can be understood 
by the following two effects.
One is the spin-dependent term of the one-gluon exchange
force (CMI). 
The other is the Pauli-blocking effects among the quarks.
As for the exotic hadrons, 
CMI has been well investigated, 
while the latter has not yet been discussed much. 
However, 
from the studies of the two-baryon interaction by the quark model,
it has been found that 
the most significant effect comes from the Pauli-blocking effects among the quarks.
This effect comes from the antisymmetrization of the quarks in the system.
Actually, the effect can be repulsive or attractive
and has strong channel dependence. We call this the quark many-body effect in the following.
Both the CMI and the quark many-body effect appear in the short-range region of the two-hadron states.
We investigate the $T_{cc}$ by using a simple quark model that includes these effects and clearly shows their roles. 

From the quark model viewpoint, when the spatial part is totally symmetric,
the $(q\bar Q)$-$(q\bar Q)$ $I(J^P)=0(1^+)$ state 
 consists of two components in the flavor-spin-color space:
 (a) $ud$ spin-0 color-${\bar 3}$ and $\bar c\bar c$ spin-1 color-${3}$ and (b) $ud$ spin-1 color-${6}$ and $\bar c\bar c$ spin-0 color-${\bar 6}$. 
For both of the components,
the color-spin interaction (CMI) is attractive, but their sizes are different;
CMI in (a) is six times more attractive than in (b). On the
other hand, the quark many-body effect in (a) is repulsive, while it is attractive in (b). 
As a result of this cancellation, we found the $T_{cc}$ as a very shallow bound state and
$T_{bb}$ as a deeply bound state.
Such a cancellation can also be seen in other exotic hadrons, such as in the $P_c$ or in the $P_{cs}$, 
where we also have near-threshold exotic states \cite{Takeuchi:2016ejt,Giachino:2022pws,Yamaguchi:2019vea}.
Our aim here is to comprehensively understand the mechanism that forms the exotic hadrons.

\section{Model}\label{sec2}

We use a simplified nonrelativistic quark model to investigate the exotic $q\bar Qq\bar Q$ mesons.
The quark and antiquark are assumed to form meson-clusters.
The quarks and the antiquarks in the system are antisymmetrized.
As for the interaction between the quarks,
we only consider the color-spin interaction explicitly.
Other terms, such as the Coulomb and the confinement potentials are assumed to be 
used to make the meson clusters, 
and the residual interaction is the color-spin interaction.
By this simplified model, one can see the quark configuration more clearly
and can discuss the situation almost free from parameter choice.

\subsection{The model space}
A conventional $q\bar Q$ meson with a quantum number $\alpha$ 
in the flavor-spin space can be written by the quark and antiquark as
\begin{align}
\phi_\alpha&=\phi_{0s}(b_{12},\vecr_{12})c^\alpha_{s_1s_2}|q(f_1,s_1)\bar Q(f_2,s_2)\ket~,
\label{eq:mesonwf}
\end{align}%
where $\phi_{0s}(b,\vecr)$ is the $0s$ harmonic oscillator wave function with a size parameter $b$,
$\vecr_{12}$ is the relative coordinate of the quark and the antiquark, 
$f_i$ and $s_i$ are the flavor and spin of the $i$th (anti)quark, respectively,
and $c^\alpha$ is the Clebsch-Gordan coefficient.
The color is taken to be singlet.
We do not use the wave function obtained from the quark Hamiltonian itself 
but use a Gaussian wave function with an appropriate size parameter as an approximation.
Using a single-range Gaussian may underestimate the quark interchange effects because it decreases faster as $r_{12}$ becomes larger compared to the solved one in the linear confinement potential. 
However, we use it here for simplicity and to clarify the situation of the four-quark states.

We estimate the size parameter for the $q\bar Q$ mesons as follows.
The observed meson mass spectrum shows that their orbital excitation energy depends not much on the flavors
 (see Table \ref{tbl:mesonmasses}). 
Suppose the potential is only the linear confinement, 
the excitation energy $\varDelta E$ is written by the reduced mass $\mu$ and the mean square radius $\bra r^2\ket$ as
\begin{align}
\varDelta E &= (a_1-a_2){c_1\over 2 \mu \bra r^2\ket} \\
&= {1\over 3}(a_1-a_2)c_1 {1\over \mu b_{12}^2} \sim 1.70 \omega_0
\label{eq:mesonsize}
\end{align}%
with $a_n$ the zeros of the Airy function ($a_1=-2.338,~a_2=-4.088,\cdots$) and a certain $c$-number $c_1$=2.9156. 
Since the Gaussian wave function in eq.\ \eqref{eq:mesonwf} gives the mean square radius ${3\over 2}b_{12}^2$,
we have eq.\ \eqref{eq:mesonsize}.
The flavor independentness of the excitation energies can be realized 
by taking the $\mu b_{12}^2$ to be a flavor-independent constant, $\omega_0^{-1}$.
The value of $\omega_0$ is about $400$ MeV
when the excitation energy $\varDelta E$ is about 680 MeV like that of the $D$ mesons.
The corresponding single particle size parameter 
for the light quark is 
about 0.57 fm. 
Whereas when we take $b$ to be 0.5 fm, the $\omega_0 \sim 500$ MeV.
We use these two values, $\omega_0 =$ 400 and 500 MeV, to examine the feature of $T_{QQ}$ in the following.

\begin{table}[th]
\caption{Masses and the excitation energies of the mesons (in MeV) \cite{Workman:2022ynf}.}\label{tbl:mesonmasses}%
\begin{tabular}{@{}lccccccccccc@{}}
\toprule
&$\omega$ & $\phi$  & $\eta_c$ & J/$\psi$,$\psi$ & $\eta_b$ & $\Upsilon$&$D$&$D^*$  \\
\midrule
$1S$  & 782.66 &1019.461 & 2983.9   & 3096.900 & 9398.7 & 9460.40 &1869.66 &2010.26 \\
$2S$  & 1410$\pm 60$& 1680$\pm 20$ &3637.7$\pm 1.1$&3686.10$\pm 0.06$ & 9999$\pm4$& 10023.4$\pm 0.5$ &2549$\pm 19$ &2627$\pm 10$\\
$\varDelta E$  & 627   & 661 & 654  &589 & 600 & 563 & 679 & 617\\
\botrule
\end{tabular}
\end{table}

As for the wave functions of the four quark systems, we take
\begin{align}
\Psi_\alpha &= (1-P^{sfc}_{24}P^{orb.}_{24})\psi_\alpha\chi_\alpha(R_{12})
\\
\psi_\alpha &= 
\sum_{\alpha_1\alpha_2}c(\alpha_1\alpha_2;\alpha)\phi_{\alpha_1}\phi_{\alpha_2}~,
\end{align}%
where $\alpha$ is the flavor and spin quantum numbers, 
and $P^{sfc}_{24}$ and $P^{orb.}_{24}$ are the quark exchange operators
between the two heavy antiquarks 
in the spin-flavor-color space and in the orbital space, respectively. 
The wave function for the relative motion between the two meson clusters, $\chi$,
is taken to be $S$-wave.
The $T_{QQ}$ is a spin-1 and isospin-0 state and has two $\psi$ components as
\begin{align}
\psi_A &= {1\over 2}(\bar D^0D^{*-}-D^{-}\bar D^{*0}+D^{*-}\bar D^0-\bar D^{*0}D^{-})\big|_{J=1}
\label{eq:psiA}
\\
\psi_B &={1\over \sqrt{2}}(\bar D^{*0}D^{*-}-D^{*-}\bar D^{*0})\big|_{J=1}
\label{eq:psiB}
\\
P^{sfc}_{24}\psi_A &={1\over 3}\psi_B,~~~P^{sfc}_{24}\psi_B={1\over 3}\psi_A~,
\end{align}%
where $\bar D^0=u\bar c$ and $D^-=d\bar c$ with the $*$ mark for the spin-1 $q\bar c$ state.
Those for the bottom system can be expressed similarly.

\subsection{Hamiltonian}

We employ the Hamiltonian for quark systems as
\begin{align}
H_q&=H_0 + V_{q\bar Q}+V_{qq,\bar Q\bar Q}
\label{eq:Hamiltonian}
\\
H_0 &= \sum_{i}(m_i+{p_i^2\over 2m_i}) - {p_G^2\over 2m_G}
\\
V_{q\bar Q}&=\sum_{i=q,j=\bar Q} \Big(
{(\lambda_i\cdot\lambda_j})a_{f_if_j}
-(\lambda_i\cdot\lambda_j)(\sigma_i\cdot\sigma_j)  c_{f_if_j}
\Big){\cal P}^{0s}
\\
V_{qq,\bar Q\bar Q}&=-\sum_{ij=qq ~or~ \bar q\bar q} (\lambda_i\cdot\lambda_j)(\sigma_i\cdot\sigma_j)  c_{f_if_j}~{\cal P}^{0s}~,
\end{align}%
where $m_i$ and $p_i$ are the mass and the momentum of the $i$th (anti)quark, and
$m_G$ and $P_G$ are the total mass and the center of mass momentum, respectively.
The potential $V_{q\bar Q}$ consists of the spin-independent color term, $(\lambda\lambda)$,
and the color-spin term, $(\lambda\lambda)(\sigma\sigma)$.
${\cal P}^{0s}$ is the projection operator onto the configuration where the meson(s) that
include the interacting quarks are
in the orbital 0s states.

The present model does not depend on the details of the orbital shape of the potentials;
$a_{f_if_j}$ and $c_{f_if_j}$ are $c$-numbers that depend on the flavors of the interacting quarks.
The color-spin term stands for the color-magnetic interaction.
The coefficients $c_{u\bar c}$ and $c_{u\bar b}$ are taken so that the term gives the observed 
$\bar D^*$-$\bar D$ and $B^*$-$B$ mass differences, respectively.
The coefficient $c_{ud}$ is obtained from the $\Lambda_c$, $\Sigma_c$ and $\Sigma_c^*$ mass difference; 
it differs only by 2.4 percent if one uses the $\Lambda$, $\Sigma$ and $\Sigma^*$ masses,
and contributes to the N-$\Delta$ mass difference by 314 MeV.
The $c_{\bar Q\bar Q}$'s are obtained from the $J/\psi$-$\eta_c$ and $\Upsilon$-$\eta_b$ mass differences
because there are not enough measurements of the doubly-charmed baryons to extract the value.
Table \ref{tbl:parameters} summarizes the parameters we use here. 
The quark masses, the kinetic term, the coefficient of the spin-independent term, $a_{u\bar Q}$, 
and the spin-dependent term, $c_{u\bar Q}$, reproduce the observed $D$ and $B$ meson mass as
\begin{align}
M_P &= m_Q+m_u + {3\over 4}\omega_0 -{16\over 3}a_{u\bar Q} + {16\over 3}\bra \sigma\sigma\ket c_{u\bar Q}~.
\end{align}%
The spin-independent color term
stands for the color-Coulomb or the confinement interaction, as well as the color-electric interaction;
it cannot be regarded as a residual interaction.
However, in the present model, the meson-meson interaction 
does not depend on the $a_{u\bar Q}\!$'s value, which only contributes to the meson mass. 
The spin-independent terms for $qq$ or $\bar Q\bar Q$ can affect the meson interaction, but they are not included in the present work.
Moreover, we only consider the interaction from gluons; 
the effects of the pion exchange will be included in future work.
\begin{table}[th]
\caption{Parameters in this model (in MeV). $C^*_{qq}={16\over 3}c_{qq}$ is the contribution of the spin-spin term to the vector mesons.}\label{tbl:parameters}%
\begin{tabular}{@{}lccccccccccc@{}}
\toprule
$m_D$ & $m_B$ & $m_u(=m_d) $ & $ m_c $ & $ m_b$ & $ C^*_{ud}$ & $ C^*_{cc}$ & $C^*_{bb}$ & $\omega_0$ \\
\midrule
1869.66  & 5279.5   & 300 & 1500 & 4700 & 104.68  & 28.25 & 15.43 & 400, 500\\
\botrule
\end{tabular}
\end{table}

\subsection{The resonating group method (RGM) equation}

By integrating the internal motion in the mesons, we have an equation of motion 
for the mesons, 
or the resonating group method (RGM) equation as
\begin{align}
\int dR' &\sum_{\beta=A,B}\Big({\cal H}_{\alpha\beta}(R,R')-E{\cal N}_{\alpha\beta}(R,R')\Big)\chi_\beta(R') = 0
\label{eq:RGMeq}\\
{\cal O}_{\alpha\beta}(R,R') &= \int d\xi d\eta d\zeta\,\psi_\alpha\delta^3(\zeta-R)\,  O_q\, (1-P_{24})\psi_\beta\delta^3(\zeta-R')~,
\end{align}%
where $O_q$ is an operator for quarks, $H_q$ or 1, and  $\alpha$ or $\beta $ corresponds to the state $A$ or $B$
in eqs.\ \eqref{eq:psiA} and \eqref{eq:psiB}.
The $\xi$, $\eta$, and $\zeta$ are the Jacobi coordinates of the four-quark system.
The normalization kernel ${\cal N}$, that corresponds to $O_q=1$, can be expanded by the $ns$-harmonic oscillator function as
\begin{align}
{\cal N}_{\alpha\beta}(R,R')&= \sum_n \phi_{ns}(b_\alpha,R)\big(\delta_{\alpha\beta}+\bar \nu_{\alpha\beta}\,z^n\big)
\phi_{ns}(b_\beta, R')
\\
{\cal N}^{orb~ex}_{\alpha\beta}(R,R')&= \sum_n \phi_{ns}(b_\alpha,R)z^n
\phi_{ns}(b_\beta, R')
\\
{\cal H}_{0~\alpha\beta}(R,R')&= (\sum_i m_i + {3\over 4}\omega_0 \times 2)\calN_{\alpha\beta} + {\cal K}_{\alpha\beta}
\\
{\cal K}_{\alpha\beta}(R,R')&= \sum_{nn'} \phi_{ns}(b_\alpha,R)K_{nn'}
\big(\delta_{\alpha\beta}+\bar \nu_{\alpha\beta}\,z^{\min(n,n')}\big)
\phi_{n's}(b_\beta, R')~,
\end{align}%
where $-\bar \nu$ is the matrix elements of the quark interchange operator in the spin-flavor-color space,
the size parameter of the relative motion between the two mesons, $b_\alpha$, and the quark mass ratio, $z$,
\begin{align}
\bar \nu_{\alpha\beta}&=-\bra P^{sfc}_{24}\ket ,~~~
b_\alpha^2 = \mu_\alpha \omega_0,~~~z = \left({m_c-m_u\over m_c+m_u}\right)^{2}\\
K_{nn}&={\omega_0\over 2}(2n+{3\over 2}),~~~K_{nn+1}={\omega_0\over 2}\sqrt{(n+1)(n+{3\over 2})}~.
\end{align}%
The reduced mass of channel $\alpha$, $\mu_\alpha$, is calculated by the quark masses, 
$\mu_\alpha=(m_u+m_Q)/2$, and $z^0=1$ for the $m_u=m_Q$ case.
For the $T_{QQ}$, the $\bar \nu$ becomes 
\begin{align}
\nu = 1+\bar \nu = \left(\begin{array}{cc}1&-{1\over 3}\\-{1\over 3}&1\end{array}\right)~.
\end{align}%

The Pauli-blocking and the quark many-body effects appear in the kinetic kernel.
Suppose we transform \eqref{eq:RGMeq} to the Schr\"odinger-type RGM equation as,
\begin{align}
\Big({\cal N}^{-1/2}{\cal H}{\cal N}^{-1/2}-E\Big){\cal N}^{1/2}\chi = 0~.
\end{align}%
The kinetic term also becomes ${\cal N}^{-1/2}{\cal K}{\cal N}^{-1/2}$.
By including the quark interchange, each of the coefficients in the kinetic term 
changes as  (for a single-channel system)
\begin{align}
K_{nn}\to K_{nn},~~~K_{nn+1}\to K_{nn+1} \sqrt{1+\bar \nu z^n\over 1+\bar \nu z^{n+1}}~.
\end{align}%
Since $0\leq z< 1$, the off-diagonal elements become larger when $\bar \nu>0$, 
which causes an attraction between the mesons in the low-energy region.
This is the many-body effect of quarks, where the final $D^{(*)}D^*$ state 
can be constructed from the initial $D^{(*)}D^*$ state with and without the quark interchange.
Later, we will show that this quark many-body effect alone gives a bound state for the $u\bar c d\bar c$ 
or $u\bar b d\bar b$ systems.
On the other hand, the system gets a repulsion when $\bar \nu < 0$,
that corresponds to the (partially) Pauli-blocking effect.
The $\bar \nu=-1$ corresponds to that the $0s^4$ state is forbidden by the Pauli-principle.
The $K_{01}$ becomes zero, and the forbidden state is expressed by an isolated $0s$ state.
It is known that there is a very strong repulsion between the hadrons in such systems.

Let us first consider the interaction between the quark and the antiquark.
Each meson made by the $q_1$ and $\bar Q_2$ and by the $q_3$ and $\bar Q_4$ is a color-singlet. 
Since we set the projection operator to the $0s$ configuration, ${\cal P}_{0s}$,  
the $V_{q\bar Q}$ can be evaluated just in the spin-flavor-color space as
\begin{align}
\bra  \alpha|&(V_{12}+ V_{34}+V_{14}+V_{32})(1-P^{sfc}_{24})| \beta\ket 
\nonumber\\
=&
\bra \alpha|(V_{12}+ V_{34})|\beta\ket -\bra \alpha|(V_{12}+ V_{34})P_{24}+P_{24}(V_{12}+ V_{34})|\beta\ket
\\
=&(M_a+M_b-M_0)|_\alpha \delta_{\alpha\beta} 
+\Big( (M_a+M_b)|_\alpha +(M_a+M_b)|_\beta-2M_0\Big) \bar \nu_{\alpha \beta} 
\\
V_{ij}&=({\lambda_i\cdot\lambda_j})\Big(a_{uc}-(\sigma_i\cdot\sigma_j) c_{uc}\Big)
\\
M_0 &= \sum_i m_i+{3\over 4}\omega_0 ~.
\end{align}%
With the quark mass and the internal kinetic term, and taking into the orbital space,
it becomes a mass term in the RGM equation with the off-diagonal term as 
\begin{align}
{\cal H}_0+{\cal V}_{q\bar q} &= 
(M_a+M_b)|_\alpha \delta_{\alpha \beta} \delta(R-R') +{\cal K}_{\alpha \beta}
\nonumber\\
&+\Big( (M_a+M_b)|_\alpha +(M_a+M_b)|_\beta-M_0\Big) \bar\nu {\cal N}^{orb~ex}_{\alpha \beta}~.
\end{align}%
As for the interaction between the $u$ and $d$ quarks, and $\bar c\bar c$ antiquarks, the kernel becomes 
\begin{align}
{\cal V}_{q q}&= V_{\alpha\beta}{\cal N}^{orb~ex} 
\\
V&={1\over 6}\left(\begin{array}{cc}C_{ud}-C_{QQ}-C_{ud}^*+C_{QQ}^* ~~& C_{ud}+C_{QQ}+C_{ud}^*+C_{QQ}^*\\
C_{ud}+C_{QQ}+C_{ud}^*+C_{QQ}^*~~&C_{ud}-C_{QQ}-C_{ud}^*+C_{QQ}^* \\\end{array}\right)
\\
C_{f_if_j} &= -{16}c_{f_if_j},~~~C^*_{f_if_j} = {16\over 3}c_{f_if_j}~.
\end{align}%

The $T_{QQ}$ has two channels, $|A\ket$ and $|B\ket$ in eqs.\ \eqref{eq:psiA} and \eqref{eq:psiB}. 
Their superposition states, $|a\ket={1\over \sqrt{2}}(|A\ket+|B\ket)$, $|b\ket={1\over \sqrt{2}}(|A\ket-B\ket)$ 
are the simultaneous eigenstates of both of the $P^{sfc}$ and the color-spin operator $-\lambda_1\lambda_3\sigma_1\sigma_3$
or $-\lambda_2\lambda_4\sigma_2\sigma_4$.
Their eigenvalues are summarized in Table \ref{tbl:eigenvalues}.
The configuration where the $ud$ quark is in the spin-0 state, $|a\ket$, gains attraction from the spin-color interaction
but repulsion from the quark many-body effect. 
The two effects cancel each other there.

\begin{table}[th]
\caption{Eigenvalues of $\nu=\bra 1-P^{sfc}\ket$ and the color-spin operator $-\lambda\lambda\sigma\sigma$.}\label{tbl:eigenvalues}%
\begin{tabular}{@{}lcccclclccccccc@{}}
\toprule
& \multicolumn{2}{c}{$ud$} & $QQ$ & \raisebox{-2mm}{$\nu$} & many-body& \raisebox{-2mm}{$-\bra \lambda\lambda\sigma\sigma|_{ud}\ket$}  & \raisebox{-2mm}{color-spin
effect}& \raisebox{-2mm}{$-\bra\lambda\lambda\sigma\sigma|_{QQ}\ket$}\vspace*{-1.5mm}\\
& color & spin & spin & & ~~effect&  & 
& \\
\midrule
$|a\ket$ & ${\bar 3}$ & 0&1 & $2\over 3$  & repulsive & $-8$  & strongly attractive & $8\over 3$
\\[1.2ex]
$|b\ket$ & $6$ & 1 & 0 &$4\over 3$ & attractive & $-{4\over 3}$  &attractive & $4$
\\
\botrule
\end{tabular}
\end{table}

\section{Results and discussion}\label{sec3}
\subsection{Quark many-body effects}
\begin{figure}[h]%
\centering
\includegraphics[width=0.6\textwidth]{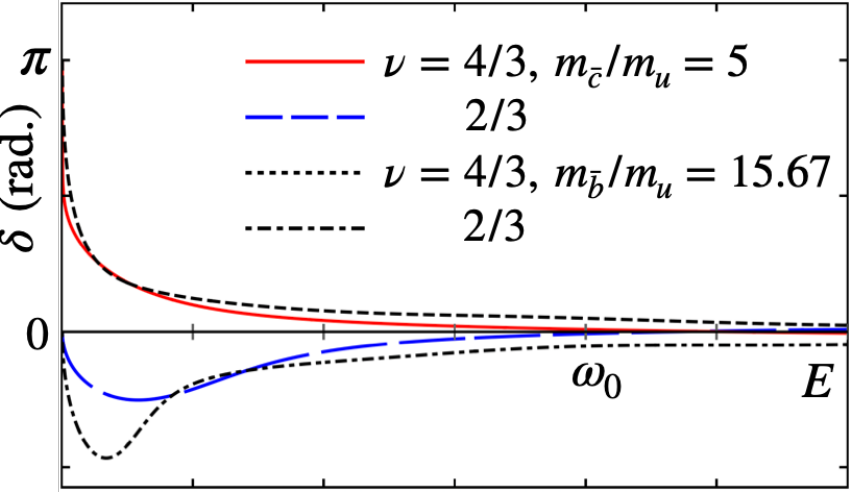}
\caption{Phase shifts obtained by the model without the interaction among quarks.
$\nu=1+\bar \nu=\bra 1-P^{sfc}\ket$. For the other notations, see text.}\label{fig:fig1}
\end{figure}

In Figure \ref{fig:fig1}, we show the phase shifts 
obtained by taking $H_q=H_0$ in eq.\ \eqref{eq:Hamiltonian}.
Since there is no color-spin interaction term, the masses of $D$ and $D^*$ are degenerated.
The short range part of the scattering states 
are the eigenstates of $P^{sfc}$, $|a\ket$ and $|b\ket$ in Table \ref{tbl:eigenvalues}, that are
the superposition of the $DD^*$ and $D^*D^*$ states.
The model has two free parameters: $\omega_0$ and $m_u/m_Q$.
For given $\nu=\bra 1-P^{sfc}\ket$, the shape of the phase shifts depend only on $m_u/m_Q$.
The $\omega_0$ gives the overall energy scale.

The quark many-body effect is not small and is something that should be taken into account.
The red solid curve in Figure \ref{fig:fig1} corresponds to the one with $\nu=4/3$ and $m_c/m_u=5$,
while the black dashed curve is for $m_b/m_u=15.67$.
Both of the systems have a bound state with a binding energy of $9.286\times 10^{-5}\omega_0$ and $1.223\times 10^{-2}\omega_0$, respectively.
The binding energy scales as $\omega_0$, and the coefficient depends on the mass ratio.
For the system with $\nu={4\over 3}$, a bound state appears at $m_{Q}/m_u > 4.6$.
The blue long-dashed and the black dot-dashed curve correspond to the phase shifts with $\nu={2\over 3}$,
both of them indicate there is a strong repulsion in this channel.

\subsection{$T_{QQ}~0(1^+)$}

The dynamical calculation with $V_{q\bar Q}$ and $V_{qq,\bar Q\bar Q}$ 
shows that there is a bound state for the $T_{cc}$ with $\omega_0=400$ MeV,
$T_{bb}$ with $\omega_0=400$ and 500MeV.
Their binding energies and the amounts of the $(0s)^4$ component, and the $ud$-spin-0 configuration, $|a\ket$,
in the  $(0s)^4$ component are summarized in Table \ref{tbl:results}.
As for $T_{cc}$, there is a very shallow bound state 
that is essentially a two-meson molecular state.
The bound state of $T_{bb}$ is a compact state.
The probability of the $ud$-spin-0 $\bar c\bar c$-spin-1 component, which has an attraction from the color-spin term,
is naturally large in $T_{bb}$.
It is found to be still a major component in $T_{cc}$ at the short range.

\begin{table}[th]
\phantom{0}
\caption{Binding energy, the size of $(0s)^4$ configuration, and $ud$ spin-1 component of $T_{QQ}$.}\label{tbl:results}%
\begin{tabular}{@{}lccccccccccc@{}}
\toprule
&$\omega_0$(MeV) & B.E.(MeV) & $(0s)^4$-prob.\ & $ud$-spin-0 prob.\ in $(0s)^4$ \\
\midrule
$T_{cc}$ &400&  \phantom{0}0.11 & 0.0392 & 0.625 \\
$T_{bb}$ &400&  18.8\phantom{0} & 0.297\phantom{0} & 0.884 \\
$T_{bb}$ &500&  \phantom{0}9.57 & 0.164\phantom{0} & 0.778 \\
\botrule
\end{tabular}
\end{table}

\begin{figure}[h]%
\centering
\includegraphics[width=0.6\textwidth]{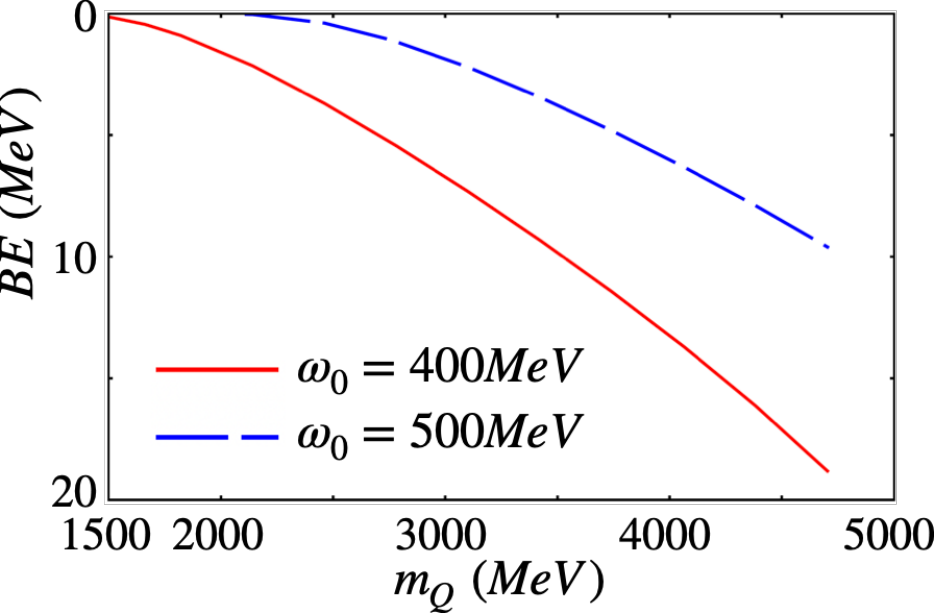}
\caption{Binding energy of the $T_{QQ}$ $I(J^P)=0(1^+)$ states with a various heavy quark mass.}\label{fig:fig2}
\end{figure}

To see the heavy quark mass dependence, 
we change the value of $m_Q$ from 1500 MeV (the charm quark) to 4700 MeV (the bottom quark).
Their binding energy is plotted in Figure \ref{fig:fig2}.
As the mass of the heavy quarks increases, the binding energy increases.
For the $\omega_0=500$ MeV case, a bound state appears for $m_Q>2111$ MeV.
The binding energy is less than 1 MeV for $m_Q<2733$ MeV.
A shallow bound state can be realized by a rather broad range parameter set.

As was pointed out in the refs.\ \cite{Karliner:2017qjm,Eichten:2017ffp,Meng:2020knc},
the color term between the heavy quarks can give a strong attraction to the $ud$ spin-0 configuration
while it gives a repulsion to the $ud$ spin-1 configuration.
It is found that with a certain parameter set, we 
can derive a situation where the $T_{cc}$ is more deeply bound, 
and the isospin-0 $B^{(*)}B^*$ channel has two bound states
like in ref.\ \cite{Meng:2023for}.
However, in the present model, we omit the term $\lambda_1\lambda_3 a_{ud}$ and $\lambda_2\lambda_4 a_{QQ}$ in the potential $V_{qq,\bar Q\bar Q}$.
It is because our ansatz here is that the size of the mesons is kept unchanged
and that the interaction is the residual one.
Since the confinement and the Coulomb potentials are proportional to $\lambda\lambda$,
they have the factor $-{8\over 3}$ in the configuration where $ud$ and $\bar Q\bar Q$ are in the color $\bar 3$-$3$.
The confinement and the Coulomb potential between the $u$ and $\bar Q$ there have the factor $-{4\over 3}$ each.
This factor is far smaller for those in the mesons, $-{16\over 3}$. 
The distance between $u$ and $\bar Q$ may differ in such states.
Introducing more configurations surely leads to more attraction and may cause a deeper bound state(s).
The simplified quark model here leaves it as an open question.

Especially for the shallow state, the one-pion exchange (OPEP) between the light quarks or the two $D$'s 
will be important. 
However, the OPEP alone cannot make $T_{cc}$ as reported in refs.\ \cite{He:2023ucd,Sakai:2023syt}.
It is necessary to include both the quark effects and the meson exchange
to discuss $T_{cc}$ as in \cite{Meng:2023for,He:2023ucd}.


\bigskip

This work is supported in part by Kakenhi Grants-in-Aid for Scientific Research [Grant No. 21H04478(A)] and  Innovative Areas (Grant No. 18H05407) (AH), Grants-in-Aid for Scientific Research [Grant No. 22H04940(S)] (MT).
This work is also  supported by the RCNP Collaboration Research Network program as the project number COREnet-2022 (project 34)  by (ST).


\bibliography{EFB25_proc_STakeuchi}


\end{document}